\documentclass{elsart}
\usepackage{graphicx}

\begin{document}

\begin{frontmatter}
\title{Correction to scaling analysis of diffusion-limited aggregation}

\author{Ell\'ak Somfai\corauthref{cor}},
\ead{e.somfai@warwick.ac.uk}
\corauth[cor]{Corresponding author.}
\author{Robin C. Ball},
\author{Neill E. Bowler} 
\address{Department of Physics, University of Warwick, Coventry,
	CV4 7AL, England}

\author{Leonard M. Sander}
\address{Michigan Center for Theoretical Physics, Department of
	Physics, University of Michigan, Ann Arbor, Michigan, 48109-1120}

\begin{abstract}
Diffusion-limited aggregation is consistent with simple scaling. However,
strong subdominant terms are present, and these can account for various
earlier claims of anomalous scaling. We show this in detail for the case of
multiscaling.
\end{abstract}


\end{frontmatter}

\section{Introduction}
\label{sec:intro}

Since its introduction by Witten and Sander in 1981 \cite{witten81},
diffusion-limited aggregation (DLA) has been the fundamental stochastic model
of quasistatic growth processes where the growth is limited by a diffusion
process.  The model can be described in simple terms: a rigid aggregate grows by
the capture of a low density of Brownian particles, which attach to it on
first contact. A highly ramified branching structure is produced, which---at
least on first sight---appears to be fractal.

One of the most basic questions asked about DLA is whether the growing
clusters obey simple scaling, i.e. are they indeed simple fractals? Based on
numerical simulations, it has been suggested that the scaling is more complex:
multiple divergent length scales might be present \cite{plischke84}, the
ensemble variance of cluster radii might have anomalous scaling
\cite{davidovitch99}, there could be more than one fractal dimension
\cite{mandelbrot02}, or the clusters might obey multiscaling where the fractal
dimension continuously depends on the position \cite{amitrano91}.

We claim that these anomalous scaling claims are \emph{wrong}, they are
misled by finite size transients. In particular, we will show in detail that
subdominant terms in the scaling account for the apparent ``multiscaling''
observed in small to medium size simulations. For clarity we should mention
that here \emph{multiscaling} refers to the space dependent fractal dimension
(anomalous scaling). This should not be confused with the well established
\emph{multifractality} of the harmonic (growth) measure, which is consistent
with the simple asymptotic scaling of the clusters.

\section{Simple scaling}
\label{sec:simple_scaling}

\begin{figure}	
\centering\includegraphics*[width=0.64\textwidth]{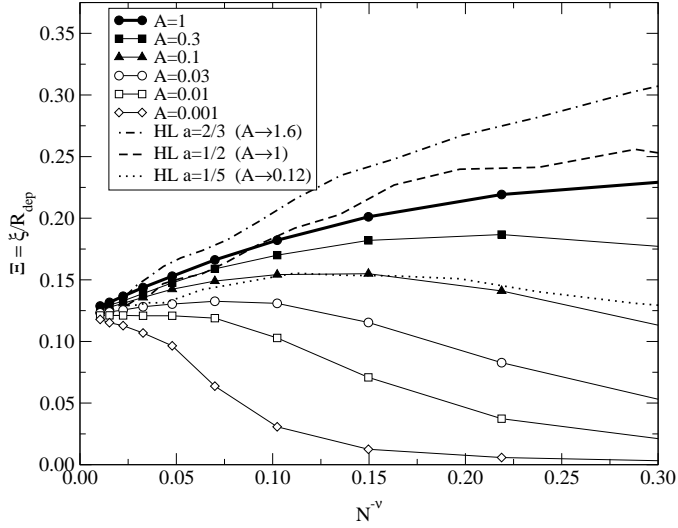}
\caption{Finite size scaling of the relative penetration depth $\Xi=\xi/
R_{\rm dep}$, with correction-to-scaling exponent $\nu=0.33$.  The thick line
connecting full circles corresponds to the standard random-walker-based DLA;
it approaches a finite asymptotic value from above. The other curves with
symbols show simulations with decreased shot noise (when a growth occurs,
instead of a full particle, only a thin layer of width $A$ is added; details
of this off-lattice noise reduction technique can be found in
Ref.~\cite{ball02pre}).  Moderate noise reduction accelerates the convergence
to the asymptotic value, while for strong noise reduction the approach is from
below. The dashed lines with no symbols correspond to simulations based on
iterative conformal maps of Hastings and Levitov \cite{hastings98}.  In all
cases the relative penetration depth approaches the \emph{same} finite
asymptotic value $\Xi_\infty=0.121\pm0.003$.}
\label{fig:penetration}
\end{figure}

In this section we will look at the scaling of various characteristic lengths:
the \emph{deposition radius} $R_{\rm dep}=\langle r \rangle$ (the average
distance of newly arriving particles from the center), the cluster's
\emph{gyration radius} $R_ {\rm gyr}=\sqrt{\frac{1}{N}\sum_{N'=1}^N\langle r^2
\rangle_{N'}}$, the \emph{root-mean-square radius} $R_2=\sqrt{\langle
r^2\rangle}$, and the \emph{penetration depth} $\xi$ (the width of the active
zone ring, where newly deposited particles land). According to our results,
DLA obeys simple scaling; all length scales scale with the same fractal
dimension. To illustrate this, we look at one of the anomalous scaling claims
mentioned in the introduction: that the cluster radius $R_{\rm dep}$ does not
scale in the same way as the penetration depth $\xi$ \cite{plischke84} ---
although it is worth mentioning that this claim has been questioned very soon
\cite{meakin85}.  The ratio of the two, often called \emph{relative
penetration depth}, \ $\Xi=\xi/R_{\rm dep}$, \  in our measurements obeys the
asymptotic form \cite{ball02pre,somfai99} for large $N$:
\begin{equation}
\Xi(N) \;\approx\; \Xi_\infty \, (1+C N^{-\nu}) \,.
\label{penetration}
\end{equation}
On Figure~\ref{fig:penetration} we plot $\Xi$ against $N^{-\nu}$ with an
appropriately chosen $\nu$; the linear behavior at $N^{-\nu}\to 0$ clearly
indicates the validity of Eq.~(\ref{penetration}). It is not easy to
obtain numerically the exponent $\nu$: systematic errors (fitting data far
from the asymptotic point) have to be balanced with large statistical errors
(fitting close to the asymptotic point). Nevertheless, all data presented in
this paper is consistent with a single ``universal'' exponent $\nu=0.33\pm
0.06$.

When the ratios of various lengths defined on DLA obey
Eq.~(\ref{penetration}), and the lengths have an asymptotic power-law
dependence on $N$, then they can be written in a scaling form with a leading
subdominant term \cite{ball02pre}:
\begin{equation}
R(N) \;\approx\; \widehat R N^{1/D} (1+\tilde R N^{-\nu})
\label{rad-fit}
\end{equation}

\begin{figure}	
\centering\includegraphics*[width=0.7\textwidth]{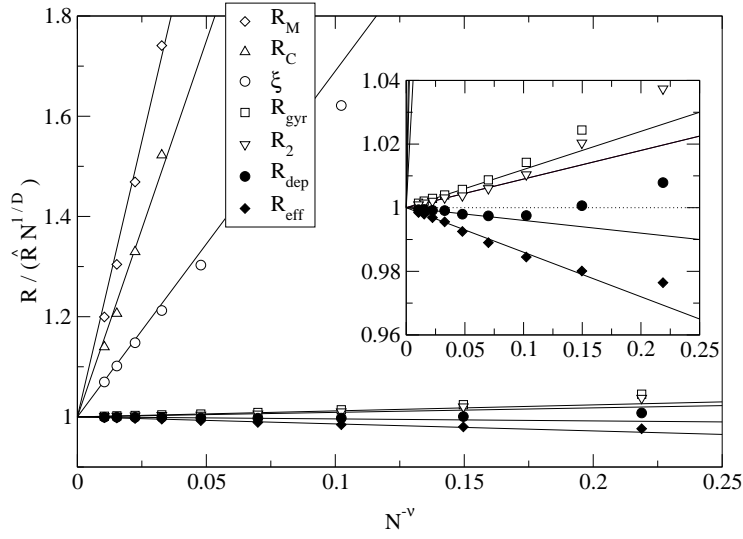}
\caption{Correction to scaling fits of various lengths, $D=1.711$ and
$\nu=0.33$. Some of the quantities have positive correction (open symbols),
others negative (filled symbols). The largest corrections is taken by lengths
having low asymptotic values.  \textbf{Inset}: $y$ axis magnified around 1.
\bigskip}
\label{fig:rad-fit}
\end{figure}

A numerical proof of this finite size scaling is plotted on
Figure~\ref{fig:rad-fit}. For completeness, we collected the corresponding
coefficients $\widehat R$ and $\tilde R$ for many characteristic lengths in
Table~\ref{tab:rad-fit}.


The coefficients are not independent, it is easy to derive some relations
between them by neglecting higher order corrections: $\widehat R_2 =
\sqrt{\widehat R_\mathrm{dep}^2 + \widehat\xi_0^2}$ and $\tilde R_2 = (\tilde
R_\mathrm{dep}\widehat R_\mathrm{dep}^2 + \tilde\xi_0 \widehat\xi_0^2) /
(\widehat
R_\mathrm{dep}^2 + \widehat\xi_0^2)$, or for the gyration radius $\widehat
R_\mathrm{gyr} = \widehat R_2/\sqrt{1+2/D}$ and $\tilde R_\mathrm{gyr} = \tilde
R_2 (1+2/D)/(1-\nu+2/D)$. The measured coefficients satisfy these relations
within error.

\section{Correction to scaling analysis of multiscaling}
\label{sec:multiscaling}

\begin{table}
\begin{center}
\begin{tabular}{ll|cc}
& definition & $\widehat R$ & $\tilde R$ \\
\hline
deposition radius & $R_\mathrm{dep} = \langle r \rangle$
	& 0.733(1) & -0.04(2) \\
root-mean-square radius & $R_2 = \sqrt{\langle r^2\rangle}$
	& 0.738(1) & 0.09(2) \\
gyration radius & $R_\mathrm{gyr} = \sqrt{\frac{1}{N}\sum_{N'=1}^N\langle
		r^2 \rangle_{N'}}$
	& 0.501(1) & 0.12(2) \\
effective (Laplacian) radius \cite{somfai99} & $R_\mathrm{eff}=\langle\exp\,
		(\int dq\ln r)\rangle$
	& 0.726(1) & -0.14(3) \\
effective radius variability & $\delta R_\mathrm{eff}=\sqrt{
		\mathop{\rm var}[\exp\, (\int dq\ln r)]}$
	& 0.0086(10) & 15 \\
maximal radius & $R_\mathrm{max}=\langle\max_q r\rangle$
	& 0.892(3) & 1.0 \\
maximal radius variability & $\delta R_\mathrm{max}=\sqrt{\mathop{\rm var}
		[\max_q r]}$
	& 0.034(2) & 13. \\
seed to center-of-charge dist. & $R_\mathrm{C} = \sqrt{\langle|\int dq\,
		{\mathbf r}|^2\rangle}$
	& 0.027(3) & 15.(10)\\
seed to center-of-mass distance & $R_\mathrm{M} = \sqrt{\langle|\frac{1}{N}
		\sum_{N'=1}^N{\mathbf r}_{N'}|^2\rangle}$
	& 0.016(1) & 22.(6)\\
ensemble penetration depth & $\xi_0=\sqrt{\langle r^2\rangle-\langle
		r\rangle^2}$
	& 0.091(1) & 6.9(8) \\
\end{tabular}
\end{center}
\medskip

\caption{Coefficients of correction to scaling fits of form
Eq.~(\ref{rad-fit}), with $D=1.711$ and $\nu=0.33$. In the definitions $r$
denotes the distance of the $N$-th particle from the seed,
$\langle\cdot\rangle$ is the average over the ensemble of clusters, and $\int
dq$ is the average over the harmonic measure of a fixed cluster. The harmonic
measure, or charge, is the probability measure of growth. The error in the
last digit (when known) is indicated in parentheses.}
\label{tab:rad-fit}
\end{table}

We start with the multiscaling assumption: it has been suggested
\cite{coniglio90} that the particle density of an
$N$-particle cluster at distance $xR_\mathrm{gyr}$ away from the center scales
with $R_\mathrm{gyr}$ with an $x$-dependent co-dimension:
\begin{equation}
g_N(xR_\mathrm{gyr}) = C(x) R_\mathrm{gyr}^{-d+D(x)}
\end{equation}
From this scaling law $D(x)$ can be obtained as
\begin{equation}
-d+D(x) = \left.\frac{\partial\ln g_{N(R_\mathrm{gyr})}(xR_\mathrm{gyr})}
    {\partial\ln R_\mathrm{gyr}}\right|_x \,.
\label{loglogder}
\end{equation}
In direct numerical measurements \cite{amitrano91} a non-trivial $D(x)$ was
obtained, forming the basis of multiscaling claims.

\begin{figure}	
\centering\includegraphics*[width=0.64\textwidth]{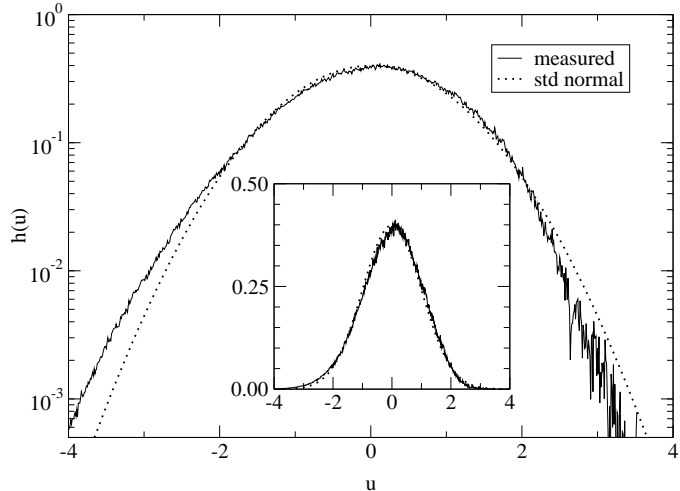}
\caption{The scaling function $h$, measured in random-walker-based simulation
(the histogram bin width is $\Delta u=0.01$), compared to standard normal
distribution. $h$ falls off faster than
Gaussian at large positive $u$, compensated with slower fall off at negative
$u$, but overall it is well approximated with the density of the standard
normal distribution. \textbf{Inset}: the same quantities on linear scale.}
\label{fig:h}
\end{figure}

Now we consider the distribution of $r$, the distance of the $N$-th particle
from the seed. From the definitions in Table~\ref{tab:rad-fit}, the mean of
$r$ is $R_\mathrm{dep}$ and its standard deviation is $\xi_0$. The probability
density of $r$ can be written as 
\begin{equation}
\frac{1}{\xi_0(N)}\,\,
      h\!\left(\frac{r-R_\mathrm{dep}(N)}{\xi_0(N)}\right) \,,
\end{equation}
if we assume that the shape $h$ of the distribution is independent of $N$.
After replacing the sum over the particles with an integral, for the particle
density we obtain
\begin{equation}
2\pi r\; g_N(r) = \int_0^N dN' \;\;
    \frac{1}{\xi_0(N')}\,\,
      h\!\left(\frac{r-R_\mathrm{dep}(N')}{\xi_0(N')}\right)
\label{our-g}
\end{equation}
A formula similar to this has been suggested earlier \cite{lee93}. At this
point we can calculate \cite{ball02pre} the function $D(x)$ from
Eqs.~(\ref{loglogder}) and (\ref{our-g}), because we already know the
correction to scaling approximation (\ref{rad-fit}) of $R_\mathrm{gyr}(N)$,
$R_\mathrm{dep}(N)$ and $\xi_0(N)$.  The only extra ingredient needed is the
functional form of $h$, which we measured directly. It is a normalized
probability density of zero mean and unit variance, and as shown on
Figure~\ref{fig:h}, it turns out to be well approximated
by the standard normal distribution.

\begin{figure}	
\centering\includegraphics*[width=0.64\textwidth]{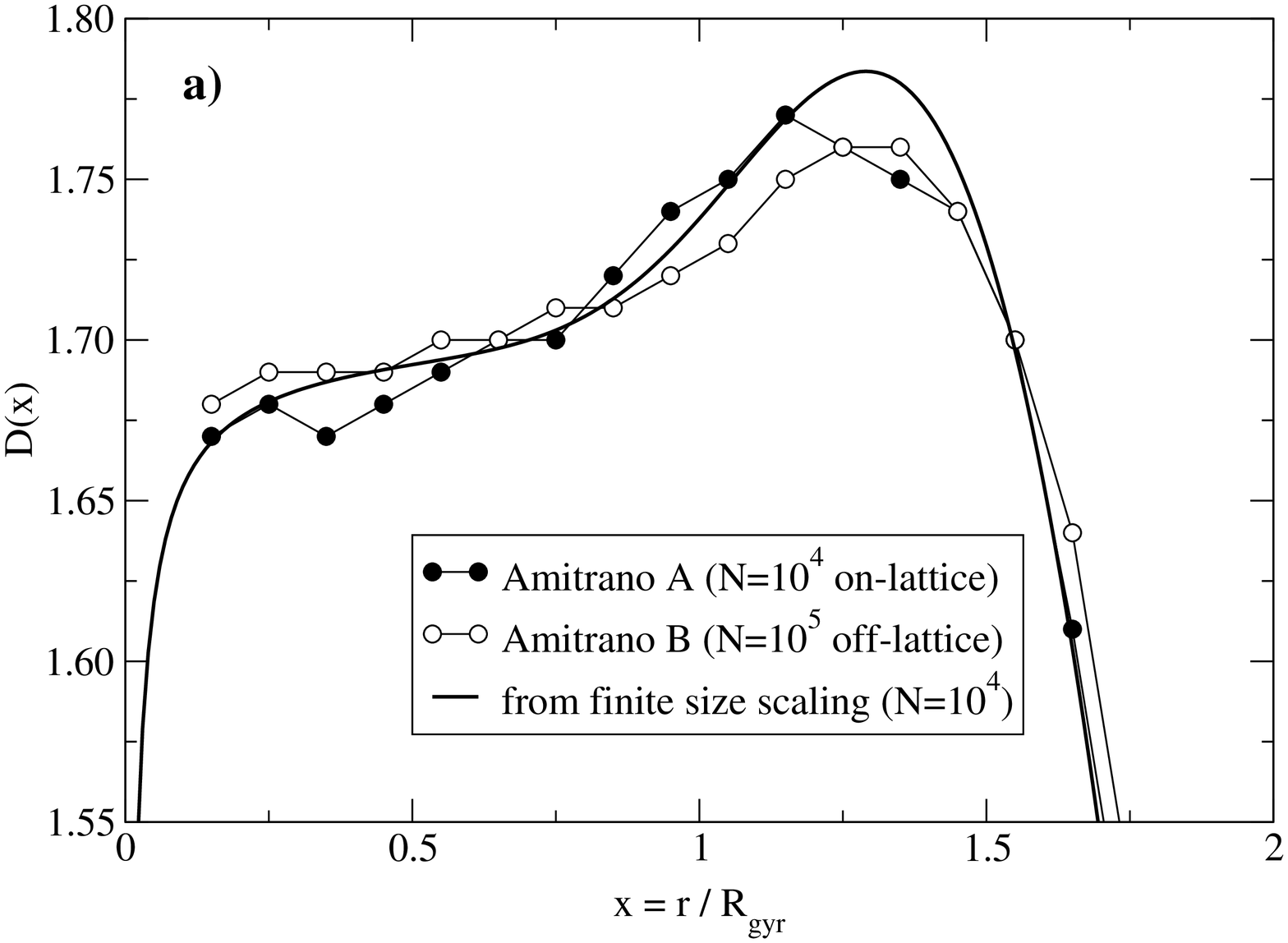}
\includegraphics*[width=0.64\textwidth]{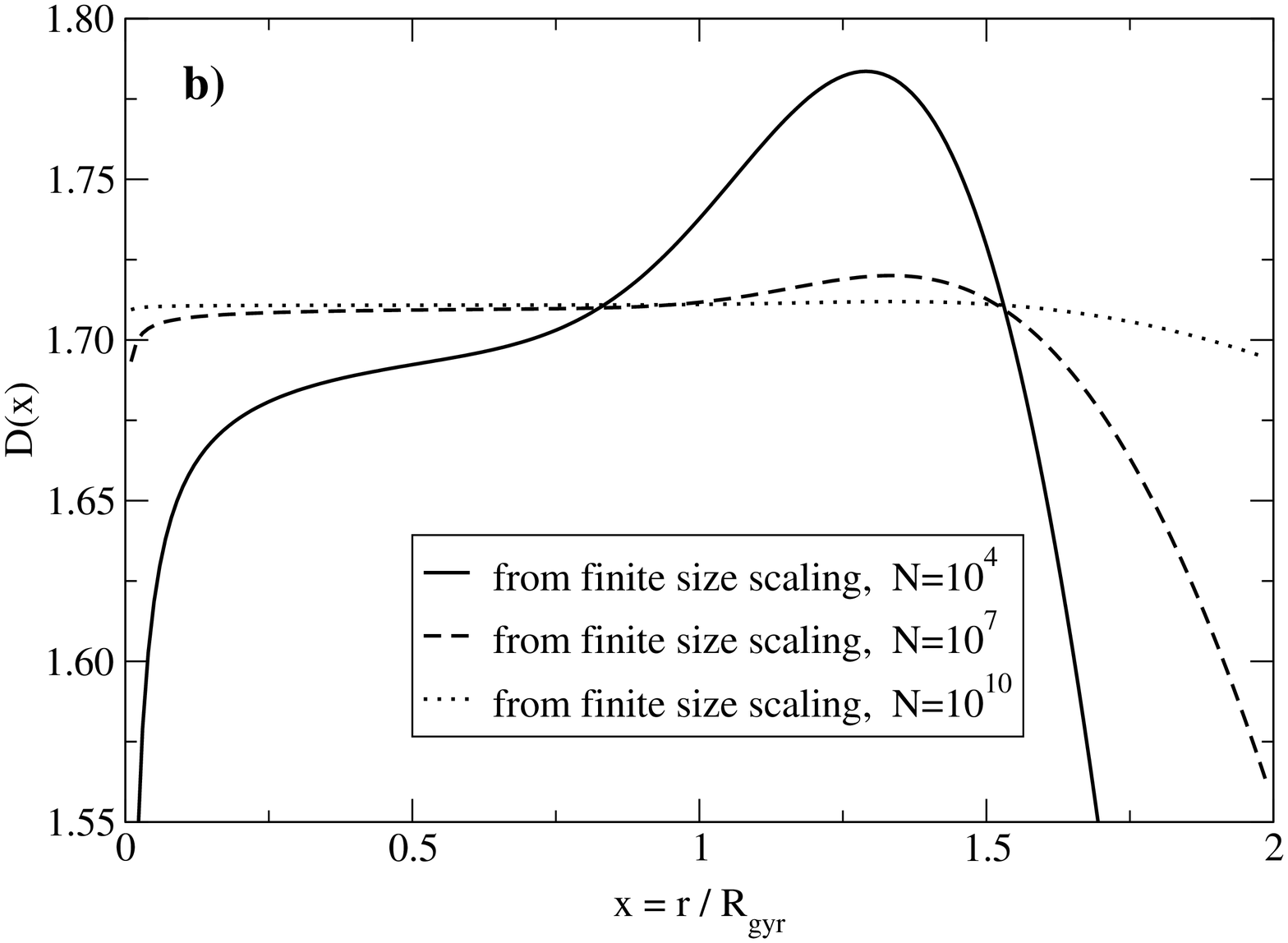}
\caption{Apparent ``multiscaling dimensions''. \textbf{a)} Comparison of the
directly measured dimensions from Ref.~\cite{amitrano91} and the finite size
scaling prediction at $N=10^4$ (the size corresponding to the simulations).
\textbf{b)} Finite size scaling predictions at sizes $N=10^4$, $10^7$ and
$10^{10}$. In the limit $N\to\infty$, the dimension approaches a constant:
$D(x)\to D$.}
\label{fig:multiscaling}
\end{figure}

Now we compare $D(x)$ calculated with correction to scaling forms with that of
earlier direct measurements on Figure~\ref{fig:multiscaling}a: the agreement
is rather good. However, our method indicates (Figure~\ref{fig:multiscaling}b)
that for larger size clusters $D(x)$ collapses to a constant: for $N\to\infty$
the radii approach pure scaling. From this we conclude that the observed
``multiscaling'' is a small size transient caused by the strong correction to
scaling of the radii (mostly of $\xi_0$).

\section{Summary}
\label{sec:summary}

We have seen that current numerical measurements are consistent with the
simple scaling picture of DLA. Subdominant terms, however, are strongly
present: earlier anomalous scaling claims---including divergent length scales
and multiscaling---were misled by them. The correction to scaling analysis,
calculating the effect of the dominant correction, explains these earlier
observations, even for a complicated quantities like $D(x)$. It remains a
challenge for the future to predict theoretically the---so far only
empirical---correction to scaling parameters.

\begin{ack}
This research has been supported by a Marie Curie Fellowship of the EC
programme ``Improving Human Potential'' under contract number
HPMF-CT-2000-00800.
\end{ack}

\bibliographystyle{elsart-num}
\bibliography{dlaref}

\end{document}